\documentclass[12pt]{article}

\newcommand{\hic}{}
\newcommand{\hhat}{}
\newcommand{\re}{{\rm e}} 
\newcommand{\vp}{\phi} 
\newcommand{\ds}{\displaystyle} 

\advance\hoffset by -7mm  
\makeatletter
\@addtoreset{equation}{section}
\makeatother
\renewcommand{\theequation}{\thesection.\arabic{equation}}

\def\ii{\'\i}

\def\ftoday{{\sl {Le \number\day \space\ifcase\month 
\or janvier\or f\'evrier\or mars\or avril\or mai
\or juin\or juillet\or ao\^ut\or septembre\or octobre
\or novembre \or d\'ecembre\fi\space \number\year}}}    
\def\ptoday{{\sl {\number\day \space de\space \ifcase\month 
\or janeiro\or fevereiro\or mar{\c c}o\or abril\or maio
\or junho\or julho\or agosto\or setembro\or outubro
\or novembro \or dezembro\fi\space de\space \number\year}}}    
\def\gtoday{{\sl {Den \number\day. \ifcase\month 
\or Januar\or Februar\or M\"arz\or April\or Mai
\or Juni\or Juli\or August\or September\or Oktober
\or November \or Dezember\fi\space \number\year}}}    
\def\today{{\sl {\ifcase\month
\or January\or February\or March\or April\or May
\or June\or July\or August\or September\or October
\or November \or December\fi \space\number\day,\space 
                                            \number\year}}}
\newcommand{\journal}[4]{{\em #1~}#2\,(#3)\,#4}

\newcommand{\ijmp}{\journal {Int. J. Mod. Phys.}}

\newcommand{\pr}{\journal {Phys. Rev.}}

\newcommand{\jmp}{\journal {J. Math. Phys.}}

\newcommand{\cmp}{\journal {Commun. Math. Phys.}}

\newcommand{\cqg}{\journal {Class. Quantum Grav.}}

\newcommand{\np}{\journal {Nucl. Phys.}}
\newcommand{\pl}{\journal {Phys. Lett.}}

\newcommand{\prep}{\journal {Phys. Rep.}}

\newcommand{\nc}{\journal {Nuovo Cimento}}

\renewcommand{\a}{\alpha}

\renewcommand{\d}{\delta}         \newcommand{\D}{\Delta}
\newcommand{\e}{\varepsilon}

\newcommand{\la}{\lambda}        
\newcommand{\m}{\mu}
\newcommand{\n}{\nu}
\newcommand{\om}{\omega}

\newcommand{\f}{{\phi}}           
\newcommand{\vf}{{\varphi}}

\renewcommand{\AA}{{\cal A}}
\newcommand{\BB}{{\cal B}}

\newcommand{\DD}{{\cal D}}

\newcommand{\FF}{{\cal F}}

\newcommand{\LL}{{\cal L}}
\newcommand{\MM}{{\cal M}}

\renewcommand{\SS}{{\cal S}}

\newcommand{\es}{\\[3mm]}

\newcommand{\sla}{\raise.15ex\hbox{$/$}\kern -.57em} 
\newcommand{\Sla}{\raise.15ex\hbox{$/$}\kern -.70em}

\newcommand{\lp}{\left(}\newcommand{\rp}{\right)}
\newcommand{\lc}{\left[}\newcommand{\rc}{\right]}

\newcommand{\complex}{{\kern .1em {\raise .47ex
\hbox {$\scriptscriptstyle |$}}
    \kern -.4em {\rm C}}}
\newcommand{\real}{{{\rm I} \kern -.19em {\rm R}}}
\newcommand{\rational}{{\kern .1em {\raise .47ex
\hbox{$\scripscriptstyle |$}}
    \kern -.35em {\rm Q}}}
\renewcommand{\natural}{{\vrule height 1.6ex width
.05em depth 0ex \kern -.35em {\rm N}}}

\newcommand{\tr}{{\rm {Tr} \,}}
\newcommand{\half}{\dfrac{1}{2}}
\newcommand{\pa}{\partial}

\newcommand{\dfud}[2]{{\displaystyle{\frac{\delta #1}{\delta #2}}}}
\newcommand{\dfrac}[2]{{\displaystyle{\frac{#1}{#2}}}}
\newcommand{\dsum}[2]{\displaystyle{\sum_{#1}^{#2}}}   
\newcommand{\dint}{\displaystyle{\int}}

\newcommand{\twiddle}{\lower.9ex\rlap{$\kern -.1em\scriptstyle\sim$}}

\newcommand{\equ}[1]{(\ref{#1})}
\newcommand{\eq}{\begin{equation}}
\newcommand{\eqn}[1]{\label{#1}\end{equation}}
\newcommand{\eea}{\end{eqnarray}}
\newcommand{\eqa}{\begin{eqnarray}}
\newcommand{\eqan}[1]{\label{#1}\end{eqnarray}}
\newcommand{\ba}{\begin{array}}
\newcommand{\ea}{\end{array}}
\newcommand{\eqac}{\begin{equation}\begin{array}{rcl}}
\newcommand{\eqacn}[1]{\end{array}\label{#1}\end{equation}}

\begin{document}


\thispagestyle{empty}

\hspace*{\fill}{{\normalsize 
\begin{tabular}{l}
{\sf PAR-LPTHE 01-65} \\
{\sf LYCEN 2001-78}  \\
{\sf UFES-DF-OP2001/2}  \\
 {\sf November 2001}\\
\end{tabular}   
 }}

\begin{center}
{{\LARGE {\bf On the Symmetries of BF Models\es
and Their Relation with Gravity}}}
\vspace{7mm}

{\large 
Clisthenis P. Constantinidis$^{*,**,1}$, 
Fran\c cois Gieres$^{***,}$\footnote{Supported in part by the 
{\it Conselho Nacional de
Desenvolvimento Cient\'\i fico e Tecnol\'ogico (CNPq -- Brazil)}.},
\es Olivier Piguet$^{**,1}$ and Marcelo S. Sarandy$^{****,1}$
}\end{center}
\vspace{1mm}

\noindent $^{*}$ {\it LPTHE, Universit\'e Pierre et Marie Curie, Boite 
126, Tour 16, 1er \'etage,
4 Place Jussieu,
F - 75252 - Paris CEDEX 05 (France).}

\noindent $^{**}$ {\it Universidade Federal do Esp\'{\i}rito Santo 
(UFES), 
CCE, Departamento de F\'{\i}sica, Campus Universit\'ario
de Goiabeiras, BR-29060-900 - Vit\'oria - ES (Brasil).}

\noindent $^{***}$ {\it 
Institut de Physique Nucl\'eaire,
Universit\'e Claude Bernard (Lyon 1), \\
43, boulevard du 11 novembre 1918,
      F - 69622 - Villeurbanne (France).}

\noindent $^{****}$ {\it Centro Brasileiro de Pesquisas F\'\i sicas 
(CBPF), 
Coordena\c c\~ao de Teoria de Campos e Part\ii culas (CCP),
Rua Dr. Xavier Sigaud, BR-150-22290-180 - Rio de Janeiro - RJ 
(Brasil).}

{\tt E-mails: clisthen@cce.ufes.br, gieres@ipnl.in2p3.fr, \\
piguet@cce.ufes.br, sarandy@cbpf.br}

\vspace{5mm}

{\small 
\noindent
{\bf Abstract}
The perturbative 
finiteness of various topological models (e.g. BF models)
has its origin in an extra symmetry of the gauge-fixed action, 
the so-called {\em vector supersymmetry}. Since 
an invariance of this type also exists for gravity
and since gravity is closely related to certain BF models, 
vector supersymmetry should also be useful 
for tackling various  aspects of quantum gravity. 
With this motivation and goal in mind,
we first extend vector 
supersymmetry of BF models to generic manifolds
by incorporating it into the BRST symmetry within the
Batalin-Vilkovisky framework.
Thereafter, we address the relationship between gravity and BF models,
in particular for three-dimensional space-time. 
}

\newpage
  
\newpage

\section{Introduction}

\setcounter{footnote}{0}
\subsection{Motivation}

As was realized in recent years, there exists a 
close relationship between BF models and gravity 
in three dimensions~\cite{horo} as well as in higher 
dimensions~\cite{freidel1}. 
(For a review of BF models~\cite{horo, bfm} and 
other topological field theories \cite{witten,ewitten}, 
see reference   \cite{birm-et-al}.) 
The perturbative finiteness of BF models can be traced back to the 
invariance of the gauge-fixed action 
under vector supersymmetry (VSUSY) 
\cite{vec-susy}-\cite{alg-ren}. 
Since a symmetry of this type  
also exists for gravity \cite{piguet}, it should impose some constraints 
on the quantum theory of gravitation and thereby prove to be useful 
for its formulation. 

Before presenting an outline of the present paper, we first 
review the results which are known for the 
three-dimensional case.

\subsubsection{Local VSUSY of the $3d$ BF model}\label{local-vsusy}

Let $G$ be a 
 matrix Lie group and let  ${\cal G}$ denote the associated Lie
algebra.
The basic variables of the $3d$ BF model
with symmetry group $G$ 
are a connection  $1$-form ${\hic A}$ and a $1$-form potential 
${\hic B}$, both taking their values in ${\cal G}$. More explicitly,  
${\hic A}={\hic A}_\m  dx^\m = {\hic A}^a_\m T_a dx^\m $ 
(similarly for ${\hic B}$) where 
the matrices $T_a$ belong to ${\cal G}$ and satisfy the relations 
\eq
\lc T_a,T_b \rc = f_{ab}{}^c T_c
\quad ,\quad 
\tr \lp  T_a T_b \rp = \d_{ab}
\ .
\eqn{Lie-alg}

The arena of the model is a smooth manifold ${\cal M} _3$ of dimension $3$. 
The action functional is given by 
\begin{equation}
\label{sbf}
S_{\rm BF}\lp {\hic A},{\hic B}\rp  = - \int_{{\cal M}_3}  \tr \lp  {\hic B}{\hic F} \rp  
\ , 
\end{equation}
 where ${\hic F} = d{\hic A} + {1 \over 2} [{\hic A},{\hic A}] = d{\hic A} + {\hic A}^2$ 
 denotes the curvature (field strength)
 of ${\hic A}$. This functional  is invariant under 
local gauge transformations of ${\hic A}$ and ${\hic B}$ parametrized
at the infinitesimal level by   ${\cal G}$-valued fields $\hic c$ and 
$\hic\phi$, respectively\footnote{Strictly speaking, 
these transformations only leave the action 
invariant if the manifold is compact without boundary 
or if the symmetry parameters vanish on the boundary~\cite{horo}.}. 
In the standard BRST framework, these parameters are turned 
into ghost fields and the symmetry transformations 
are described by the BRST operator $s$. 
All fields are then characterized 
by a {\em total degree} which is the sum of their form degree and their 
ghost-number and all commutators are assumed to be graded with respect
to this degree\footnote{See appendix for further details.}. 
The $s$-variations of the basic fields and ghosts read as
\begin{eqnarray}
   s {\hic A} \!\!\! & = & \!\!\! - D_{{\hic A}} {\hic c}
  \qquad \quad \quad \ \ , \quad 
  s {\hic c} = - {\hic c}^2 
  \nonumber \es
    s {\hic B}  \!\!\! & = & \!\!\!  - D_{\hic A} \hic\phi - [{\hic c},{\hic B}]
  \quad , \quad 
  s \phi = - [{\hic c}, \hic\phi ]     
    \ , 
    \label{trabf}
\end{eqnarray}
where $ D_{{\hic A}} {\hic c} \equiv d{\hic c} + [ {\hic A} ,{\hic c} ]$
denotes the Yang-Mills covariant derivative. 

After performing the gauge-fixing in a Landau-type 
gauge~\cite{Maggiore-Sorella, lps, alg-ren, bvwien} 
(or in non-covariant versions of the latter~\cite{non-cov-gauges}), 
the gauge-fixed action is  invariant under VSUSY-transformations. 
At the infinitesimal level,  these transformations are 
described by an operator $\d_{\tau}$  
where $\tau \equiv \tau^\mu \partial_{\mu}$ 
is a $s$-invariant vector field of ghost-number
zero. The variation $\d_{\tau}$ 
acts as an antiderivation 
(odd operator) which lowers
the ghost-number by one unit and which anticommutes with $d$. 
Its action on the ghost fields $\hic c$ and $\hic\phi$ 
is given by~\cite{Maggiore-Sorella}
\begin{equation}
   \label{vsbf}
   \d_{\tau} {\hic c} = i _{\tau} {\hic A} 
   \quad , \quad  
 \d_{\tau} \phi = i _{\tau} {\hic B}
 \, ,
\end{equation}
where $i _{\tau}$ denotes the interior product with respect to 
the vector field $\tau$ (see appendix for technical details concerning operators,
vector fields and differential forms).
The operators $s$ and $\d_{\tau}$  satisfy a   
graded algebra of Wess-Zumino type,  
\begin{equation}
\label{algebra}
[ s, \d_{\tau} ] \, = \, \LL_{\tau} \, + \, \mbox{equations of motion}
\ ,
\end{equation}
where $\LL_{\tau}$ denotes the Lie derivative (see \equ{Lie-deriv}) 
with respect to the vector field $\tau$. 

To be more precise, we should note two points. 
First, 
invariance under VSUSY 
was shown to exist solely on manifolds
admitting vector fields that are covariantly constant with respect to some 
background metric, the vector field $\tau$
being any one of these~\cite{cs-lucch-pig}. 
VSUSY  therefore represents a {\em rigid} symmetry.
This restriction has its origin in a 
particular way of implementing the
gauge-fixing and introducing VSUSY-transformations. 
As we shall see in the present paper, VSUSY can hold 
as a truly {\it local} symmetry on a generic manifold 
if it is implemented in a different way.

Second, VSUSY only holds as an exact invariance as long as one does 
not incorporate external sources (coupling to the non-linear $s$-variations
of the fields). 
Once the latter are included into the action, 
VSUSY is expressed by a {\em broken Ward identity}, 
the breaking term being linear in the quantum fields and thus unproblematic
in quantum theory.

\subsubsection{$3d$ gravity}

The basic variables of $3d$ gravity are the dreibein 
$1$-forms $(e^i)_{i=1,2,3}$ and the Lorentz connection 
$1$-form $\omega = (\omega_{ij})_{i,j = 1,2,3}$ which takes its values 
in the Lie algebra $so(3)$ or $so(1,2)$, 
i.e. $\omega_{ij}=-\omega_{ji}$.
The action functional 
\cite{peldan} 
\begin{equation}
\label{sgrav}
S_{\rm grav}\lp e,\omega\rp  = - \int_{{\cal M}_3} \varepsilon_{ijk} e^i R^{jk}
\ , 
\end{equation}
 in which $R = d \omega + \omega^2$ denotes the curvature $2$-form, 
 is invariant under diffeomorphisms (general coordinate transformations)
 and under local Lorentz transformations. At the infinitesimal level, 
 these symmetries are parametrized, respectively, by 
a vector field $\xi = \xi^{\mu} \pa_{\mu}$ 
and an antisymmetric matrix $\Omega$. 
In the BRST formalism, the latter parameters represent ghost fields and the 
$s$-variations have the form~\cite{exp-xi}
\eq\ba{lcl}
 \label{trgr}
 s \omega  = - D_{\omega} \Omega  + \LL_{\xi} \omega 
   &\ , \ &
s \Omega = - \Omega^2 +  \LL_{\xi} \Omega 
\es
s e =   - \Omega e   +   \LL_{\xi} e 
   &\ , \ &
   s \xi = \ds{1 \over 2} \, {[ \xi , \xi ]}   
   \ .  
\ea\eqn{brst-var0}
Here, $D_{\omega} \Omega = d \Omega + [ \omega ,\Omega ]$
denotes the Lorentz 
covariant derivative, $\LL_{\xi} = i_{\xi} d-  d i_{\xi}$ represents 
the Lie derivative with respect to 
the ghost vector field $\xi$
and 
the vector field $[ \xi , \xi ]$
is the graded Lie bracket of $\xi$ with itself: 
$[ \xi , \xi ]^{\mu} =  2\xi^{\nu} \pa_{\nu} \xi ^{\mu}$.

 The action (\ref{sgrav})
can be gauge-fixed by considering a background metric 
and by choosing the 
Landau gauge. The gauge-fixed action is then invariant under 
VSUSY-transformations parametrized by a Killing vector field 
$\tau = \tau^{\mu} \pa_{\mu}$~\cite{piguet}: the latter only act
non-trivially on the ghost field $\xi$ 
according to 
\begin{equation}
   \label{vsgr}
   \d_{\tau} \xi = \tau 
\end{equation}
 and they satisfy the algebra \equ{algebra}.

\subsubsection{Relation between $3d$ gravity and BF models}\label{grav-bf}

We now describe 
the correspondence between $3d$ gravity and 
BF models~\cite{horo} 
(see also~\cite{ezawa}). As symmetry group of the BF model, one chooses 
$G= SO(3)$ or  $G= SO(1,2)$ as in the last subsection. Then, the connection 
$1$-form ${\hic A} = ({\hic A}_{ij})_{i,j = 1,2,3}$ and the $1$-form potential 
${\hic B} = ({\hic B}_{ij})_{i,j = 1,2,3}$ both represent antisymmetric matrices 
of $1$-forms. 
The correspondence between the degrees of freedom involved 
in both theories can be made more precise 
by writing 
\eq\ba{ll}
   {\hic B}_{jk}   =  \varepsilon_{ijk} e^i 
  \quad   , \quad  
&
{\hic A}_{jk} = \omega_{jk} \quad  
  ({\rm hence} \ \, {\hic F}_{jk}=R_{jk}) 
  \es
{\hic\phi}_{jk} =  i_{\xi} {\hic B}_{jk}   \quad , \quad
&{\hic c} = \Omega + i_{\xi} \omega  \ ,
\ea\eqn{change}
where $\varepsilon_{ijk}$ denotes the components of the totally antisymmetric 
tensor. 
The action (\ref{sgrav}) then goes over into the action (\ref{sbf}): 
\begin{equation}
\label{sbf1}
S_{\rm {\hic B}{\hic F}}\lp {\hic A},{\hic B}\rp  = 
 - \ds{1 \over 2} 
 \int_{{\cal M}_3} {\hic B}_{jk} {\hic F}^{jk}
\ . 
\end{equation}
Furthermore, the VSUSY-variations (\ref{vsgr}) imply the variations
\equ{vsbf}.  The transformation laws of $\omega$ and $e^i$
as given by \equ{brst-var0} become 
\[\ba{l}
 s {\hic A}  =  - D_{{\hic A}} {\hic c}  + i_{\xi}{\hic F}  \es
    s {\hic B}  =  - D_{{\hic A}} \phi - [ {\hic c} , {\hic B}] 
+   i_{\xi}  D_{{\hic A}}{\hic B}   \ .
\ea\]
If the equations of motion 
${\hic F} =0 = D_{{\hic A}}{\hic B}$ 
of the BF model
are taken into account, the latter transformation laws coincide with 
those given in eqs.(\ref{trabf}). 
In summary, the actions of both models 
coincide exactly and the symmetry
transformations coincide on-shell. 

Some comments concerning these results 
are in order. First, we note that 
the reparametrization $({\hic c}, {\hic\phi}) \to (\Omega , \xi)$ 
considered in eqs.(\ref{change})
represents a field-dependent change of the generators
of the BRST differential algebra:  
this explains the appearance of the equations of motion
upon passage from one model to the other one. 
Second, we remark that there is a problem of invertibility 
 with the change of generators 
 $\xi^{\mu} \to {\hic\vp} = i_{\xi} {\hic B} \equiv \xi^{\mu} {\hic B}_{\mu}$ 
since one cannot express $\xi^{\mu}$ in terms of ${\hic\vp}$,  unless
the 3-bein $e_\m^i$ = $\frac{1}{2}\e^i{}_{jk}{\hic B}_\m^{jk}$ 
represents a nonsingular matrix in whole space-time.  
As a matter of
fact,   
this invertibility problem 
reappears 
in the perturbative approach
to quantum theory, since 
the $BF$ model is perturbed around the configuration ${\hic B}=0$
whereas the corresponding configuration in gravity, i.e.  
$e^i =0$,  represents a singular metric.  
(For a general discussion, see references~\cite{mat}
and the remarks made in~\cite{ewitten}.)

\subsection{Program} 

The results we just summarized suggest the following line of
investigation:
\begin{enumerate}
\item to generalize the results concerning VSUSY of BF models 
(in three and, more generally, in higher
dimensions)  to {\em generic} manifolds, not just those admitting 
covariantly constant vector fields,
\item to promote the three-dimensional on-shell
results  to  {\em off-shell} results,
\item to extend this correspondence
to  the higher-dimensional case and to exploit its consequences. 
\end{enumerate}

In the present paper, we will discuss VSUSY 
on generic $3$- and $4$-manifolds 
for BF models  which, in addition, involve a 
cosmological term in their action.  
(The higher-dimensional case is analogous to the $4$-dimensional one
in that it involves the phenomenon of ``ghosts for ghosts" which 
does not occur in the $3$-dimensional case.)
Moreover, some results concerning the relationship 
between gravity and BF models will be 
presented. 

At different stages of the discussion, the equations of motion of the models
appear in the transformation laws of fields 
and therefore we will follow the 
Batalin-Vilkovisky (BV) approach~\cite{bat-vilk} 
to the description of symmetries. 
Henceforth, antifields are to be introduced into the formalism
from the beginning on.
They play the role of 
external sources coupled to the BRST-variations of the basic fields.
In a final step, the antifields are redefined according to the 
Batalin-Vilkovisky prescription in order to implement
the gauge-fixing. This procedure of 
gauge-fixing does not to interfere with vector
supersymmetry thanks to the fact that 
the latter is incorporated directly into the
BRST operator.

\section{3d BF model with cosmological term}\label{BF-model-3d}

   \subsection{The model and its symmetries}\label{the-model}

To start with, we consider an arbitrary 
gauge group which will be specialized to $SO(3)$ or $SO(1,2)$ when
discussing gravity. 
The notation is the one introduced in subsection \ref{local-vsusy}.

The  action for a BF model with  ``cosmological constant''
$\alpha$ on a  $3$-manifold $\MM_3$ reads
\begin{equation}
S_{\rm inv} \lp {\hic A},{\hic B}\rp  
= - \int_{\MM_3} \tr \lp  {\hic B}{\hic F} 
+ \frac{\alpha}{3} {\hic B}^3 \rp 
\, , 
\label{bf3d}
\end{equation}
where $\alpha$ represents a real dimensionless parameter.
 The equations of motion of this model are given by 
\begin{equation}
{\hic F} + \alpha {\hic B}^2 = 0 \quad  , \quad {\hic D}{\hic B} = 0 \ ,
\label{eom3d}
\end{equation}
where ${\hic D}$ denotes the covariant derivative:
${\hic D}\cdot  = d\cdot + [{\hic A},\cdot]$.  

The symmetries of the action (\ref{bf3d})
can be expressed  in terms of 
{\em horizontality conditions}~\cite{horizon, bertlmann} 
which have the form 
\begin{eqnarray}
\tilde F  \!\!\!&=&\!\!\! 
{\hic F} - \alpha { [ {\hic B} , {\hic\vp} ] } - \alpha {\hic\vp}^2
\nonumber \es 
\tilde D \tilde B \!\!\!&=&\!\!\!  {\hic D}{\hic B}
\, , 
\label{horizon3d}
\end{eqnarray}
where the ``tilded" quantities are defined by 
\begin{eqnarray}
\tilde{A}  \!\!\!  &\equiv &\!\!\!  {\hic A} + {\hic c}
\quad  , \quad 
\ \tilde{F}  \equiv \tilde{d} \tilde{A} 
+ {\tilde{A}}^2
\quad , \quad
{\tilde{d} \equiv d + s 
} 
\label{deftildes1}
\es
\tilde{B} \!\!\! & \equiv & \!\!\!  {\hic B} + {\hic\vp}
\quad , \quad 
\tilde{D} \cdot  \equiv \tilde{d} \cdot  + [\tilde{A} , \cdot \, ]	
\ .
\nonumber 
\end{eqnarray}
Relations (\ref{horizon3d}) yield the BRST 
transformations
\begin{eqnarray}
s {\hic A} \!\!\!&=&\!\!\! - {\hic D} {\hic c} - \alpha [{\hic\vp} ,{\hic B} ] \nonumber \es
s {\hic B} \!\!\!&=&\!\!\! - [{\hic c}, {\hic B}] - {\hic D} {\hic\vp}  \nonumber \es
s {\hic c} \!\!\!&=&\!\!\! - {\hic c}^2 - \alpha {\hic\vp}^2
\nonumber \es
s {\hic\vp} \!\!\!&=&\!\!\! -[{\hic c}, {\hic\vp} ]
\ ,
\label{sym3d}
\end{eqnarray}
where the ghosts ${\hic c}$ and ${\hic\vp}$ parametrize local gauge transformations 
of the potentials ${\hic A}$ and ${\hic B}$, respectively.  

Following the lines of the BV-formalism, 
we now include antifields  ${\hic A}^*, {\hic B}^*$ and ${\hic c}^*, {\hic\vp}^*$ 
associated to the basic fields ${\hic A},{\hic B}$ and to the ghosts ${\hic c}, {\hic\vp}$. 
It is convenient to introduce the complete ladders (generalized fields or 
extended forms~\cite{ikemori})
\begin{eqnarray}
{\hic\AA} \!\!\!&=&\!\!\! \hic c + {\hic A} + {\hic B}^* + {\hic\vp}^* 
\nonumber \es
{\hic\BB} \!\!\!&=&\!\!\! {\hic\vp} + {\hic B} + {\hic A}^* + {\hic c}^* 
\, ,
\label{antifields3d}
\end{eqnarray}
as well as the extended differential
\eq
{\hic \d}=d+s
\ .
\eqn{ext-diff}
As usual, the exterior derivative $d$ is assumed to 
anticommute with the BRST operator\footnote{Unlike the BRST differential
considered up to now, the operator $s$
introduced in \equ{ext-diff}
will act on the antifields as well. It will turn out to
be the ``linearized Slavnov-Taylor operator'' which will be defined later on. 
For simplicity, we shall keep the notation $s$ for this operator
and refer to the corresponding 
variations of fields and antifields as ``BRST-transformations''.}
$s$.
The generalized field strengths associated to $\AA$ and $\BB$ 
are defined by 
\eq
{\hic\FF} \equiv {\hic\d} {\hic\AA}  +  {\hic\AA}^2   \quad , \quad 
{\hic\DD} {\hic\BB} \equiv {\hic\d} {\hic\BB} +[{\hic\AA}, {\hic\BB} ]	\ .
\eqn{gen-field-str}
The action of the BRST operator $s$ 
 is again defined in terms of horizontality
conditions, namely the ``zero curvature'' 
conditions~\cite{ikemori, carvalho, bvwien}
\begin{equation}
{\hic\FF} + \a {\hic\BB} ^2 = 0
  \quad  , \quad \DD {\hic\BB} = 0\ ,
\label{gradeeq}
\end{equation}
which imply the {\it nilpotency} of the extended operator ${\hic\d}$
and thus of $s$.
 These horizontality conditions 
have the same form as the equations of motion (\ref{eom3d}),
with ${\hic A}$ and ${\hic B}$ substituted by ${\hic\AA}$ and ${\hic\BB}$, 
and they are equivalent to\footnote{We note that, up to field redefinitions, 
these equations are 
the most general ones which are compatible with
conservation of the total degree,  
provided one imposes the discrete symmetry 
${\hic\AA} \to {\hic\AA} \, ,  {\hic\BB} \to -{\hic\BB}$.} 
\begin{equation}
s  {\hic \AA} = - ( d{\hic\AA} + {\hic\AA}^2 +  \a {\hic\BB} ^2 )
\quad , \quad 
s {\hic\BB} =  - ( d {\hic\BB} + [ {\hic\AA} , {\hic\BB}] )
\ . 
\label{svar}
\end{equation}
The action of the nilpotent BRST operator $s$ 
 on the fields and antifields is found by 
expanding relations (\ref{svar}) with respect to the ghost-number. 
In doing so, we obtain 
\eq\ba{l}
s{\hic c} =   - {\hic c}^2 - \a {\hic\vp}^2 \es
s{\hic A} = - {\hic D} {\hic c} - \a \lc {\hic\vp}, {\hic B} \rc   \es
s{\hic B}^* = - ({\hic F} + \a {\hic B}^2 )
- \lc {\hic c}, {\hic B}^* \rc - \a \lc {\hic\vp} , {\hic A}^* \rc    \es
s{\hic\vp}^* = - {\hic D}{\hic B}^* - \lc {\hic c}, {\hic\vp}^* \rc - \a 
\lp  \lc {\hic\vp} , {\hic c}^* \rc + \lc {\hic B}, {\hic A}^* \rc  \rp  
\ea\eqn{brsaf1}  
and          
\eq\ba{l}
s{\hic\vp} = - \lc {\hic c}, {\hic\vp} \rc  \es
s{\hic B} = - {\hic D} {\hic\vp} - \lc {\hic c}, {\hic B} \rc \es 
s{\hic A}^* = - {\hic D}{\hic B} - \lc {\hic c}, {\hic A}^* \rc - \lc {\hic B}^*, {\hic\vp} \rc \es 
s{\hic c}^* = 
- {\hic D}{\hic A}^* -\lc {\hic c}, {\hic c}^* \rc
 -\lc {\hic B}^*, {\hic B} \rc -\lc {\hic\vp}^*, {\hic\vp} \rc  
\ .
\ea\eqn{brsaf2}  
 
If all antifields are set to zero, 
we recover the BRST-transformations \equ{sym3d} which have been deduced
from
 the horizontality conditions (\ref{horizon3d}). However, 
in addition, we also get
the field equations \equ{eom3d}.

The BRST operator $s$ defined by \equ{gradeeq} or 
by \equ{brsaf1} and \equ{brsaf2} can be interpreted as the 
``linearized Slavnov-Taylor operator''~\cite{alg-ren}
$\SS_S$ associated to a certain action 
$S({\hic A},{\hic B},{\hic c},\f,{\hic A}^*,{\hic B}^*,{\hic c}^*,
\f^*)$: the latter operator has the form 
\eq\ba{l}
\\
\SS_S X \equiv (S,X) =
  \dsum{\vf}{}\dint
 \lp \dfud{S}{\vf^*}\dfud{X}{\vf} + 
  \dfud{S}{\vf}\dfud{X}{\vf^*} \rp 
  \qquad   {\rm with} \ \;  
\vf \in 
    \left\{  {\hic A},\, {\hic B},\, {\hic c} ,\, {\hic\vp} \right\}
\ ,
\ea\eqn{slavnov-lin}
 where $(\cdot,\cdot)$ denotes the Batalin-Vilkovisky (BV) bracket 
whose properties are depicted in the appendix, see eq.\equ{BV-bracket}. 
Indeed, given the $s$-variations \equ{brsaf1} and \equ{brsaf2}, 
we can find an action $S$ solving the functional 
differential equations  (see \equ{Bf-vphi})
\eq\ba{l}
\SS_S\vf 

\equiv
\dfud{S}{\vf^*} = s\vf 
  \quad , \quad 
\SS_S\vf^* 
\equiv
\dfud{S}{\vf} = s\vf^*
\ .
\ea\eqn{differential-eqs}
 The solution is given by the $BF$-like action
\eq
S = - \int_{\MM_3} \tr \left. \lp  {\hic\BB} ( d{\hic\AA}+{\hic\AA}^2 )
+ \dfrac{\a}{3}{\hic\BB}^3  \rp  
\right|_3 \ , 
\eqn{action}
where 
the integral is performed over all contributions of form 
degree 3~\cite{ikemori}. 
When expanded into components, 
this action takes the familiar form
\eq
S = - \int_{\MM_3} \tr \lp  {\hic B} {\hic F}  + \dfrac{\a}{3} {\hic B}^3 \rp 
  + \dsum {\vf}{} \int_{\MM_3} \tr \lp  \vf^* s\vf \rp  
  \equiv S_{\rm inv} ( {\hic A},{\hic B} ) + 
S_{\rm antifields} ( \vf,\vf^* )
\, .
\eqn{action-comp}
Moreover, the action $S$ which solves the differential equations
\equ{differential-eqs} obeys  
the (nonlinear) {\it Slavnov-Taylor identity}  or BV {\it master
equation}
\eq
\ba{l}\\
\SS(S) \equiv \half (S,S) = \dsum{\vf}{}\dint
  \dfud{S}{\vf^*}\dfud{S}{\vf}  = 0 \ .
\\ \ea
\eqn{slavnov} 
This follows from the nilpotency of $s$, which implies the nilpotency of
the operator $\SS_S$ defined by 
 \equ{slavnov-lin} and \equ{differential-eqs}, and from
the following identity that results from  \equ{special-jacobi}:
\eq 
 \SS_X \SS(S) + (\SS_S)^2 X =0 
 \qquad 
  \mbox{for any functional} \  X(\vf,\vf^*)
 \ .
 \eqn{jacobi}
Indeed, by applying \equ{jacobi} to $X$ = $\vf$ or $X=\vf^*$ 
 and by using $\SS_S{}^2=0$, we find
\eq
\dfud{\SS(S)}{\vf^*} = 0 = \dfud{\SS(S)}{\vf} 
\ ,
\eqn{proof-slavnov}
i.e. $\SS(S)$ does not depend on the fields and
antifields, hence it vanishes.

\subsection{Including diffeomorphisms and VSUSY}\label{diff+vsusy}

\subsubsection{Diffeomorphisms as 
   dynamical symmetry or as
external symmetry} 

Diffeomorphism (or general coordinate) invariance is 
already present, as a dynamical symmetry of the model,  
in the gauge invariances as 
defined by the BRST-transformations \equ{sym3d}
or \equ{brsaf1}\equ{brsaf2}. 
This can be seen~\cite{horo, ezawa} by rewriting the ghosts 
$c$ and $\vp$ as  (c.f. \equ{change})
\eq
{\hic c} = i_{\xi}  {\hic A}
\quad  ,\quad 
{\hic\vp} = i_{\xi} {\hic B}
\ ,
\eqn{iwA-B}   
where $\xi$ denotes a vector field of ghost-number one.
In fact, by substituting expressions \equ{iwA-B} 
in the gauge transformations of 
${\hic A}$ and ${\hic B}$ as given by \equ{sym3d}, we obtain the infinitesimal
transformations 
\eq\ba{l}
\d_{\xi} {\hic A} = \LL_{\xi} {\hic A} -i_{\xi} \lp  {\hic F}+\a {\hic B}^2 \rp \es
\d_{\xi} {\hic B} = \LL_{\xi} {\hic B} -i_{\xi} {\hic D}{\hic B}
\ . 
\ea\eqn{diff-dyn}
Up to terms involving 
equations of motion, these variations represent
the action of diffeomorphisms generated by the vector field $\xi$.
This shows that diffeomorphisms constitute a subgroup of 
the group of gauge transformations of the theory.

Nevertheless, it turns out to be useful to 
introduce diffeomorphisms in an independent and explicit way, i.e. as 
``external'' (non-dynamical) symmetries generated by the  
vector field\footnote{We note that our procedure is reminiscent of 
four-dimensional topological Yang-Mills theories \cite{ouvry}  
where the generic shifts of the gauge field ${\hic A}$ 
given by ${\hic \d} {\hic A} = \psi$ involve, as a special 
case, infinitesimal gauge transformations corresponding to 
$\psi = -{\hic D}c$. 
Nevertheless, both the shift and gauge 
symmetries are included into the BRST operator.} 
$\xi$. They will act by virtue of the Lie derivative $\LL_\xi$ 
on all fields including 
ghosts and antifields.
The ghost vector field $\xi$ itself transforms non-trivially 
under BRST-variations, 
but to start with,  
its transformation law will {\em not} be specified.

\subsubsection{Vector supersymmetry} 
   
 Let us recall some facts concerning vector 
supersymmetry~\cite{lps, alg-ren, bvwien} viewed
as an extra symmetry besides 
BRST invariance.
The VSUSY-transformations parametrized by a
vector field $\tau$ 
are given by
\cite{bvwien}
\begin{equation}
\d_{\tau}{\hic\AA} = i_{\tau}{\hic\AA} 
\quad  , \quad 
\d_{\tau}{\hic\BB} = i_{\tau}{\hic\BB}
\ , 
\label{vsusy1}
\end{equation}
which relations are equivalent to 
\eq\ba{lcl}
   \d_{\tau}{\hic c}  =  i_{\tau}{\hic A} 
    & \quad  , \quad
&
 \d_{\tau}{\hic\vp}  =  i_{\tau}{\hic B} 
\es
\d_{\tau}{\hic A} = i_{\tau}{\hic B}^* 
   & \quad  , \quad
&
\d_{\tau}{\hic B} = i_{\tau}{\hic A}^*  
  \es
\d_{\tau}{\hic B}^* = i_{\tau} {\hic\vp}^*  
    & \quad  , \quad
&
\d_{\tau}{\hic A}^* = i_{\tau} {\hic c}^* 
  \es
\d_{\tau} {\hic\vp}^* =0 
   & \quad  , \quad
&
\d_{\tau} {\hic c}^* =0
\ . 
\ea\eqn{vsusy2}
These variations represent a generalization (involving antifields) 
of the transformation laws \equ{vsbf}. From eqs.\equ{svar}\equ{vsusy1} 
and the
commutation relations $[d, \d_{\tau} ] = 0 = [s, i_{\tau}]$,  
 it follows that 
 the algebra $\left[ \d_{\tau}, s \right]= {\cal L}_{\tau}$
is satisfied on the extended forms ${\hic\AA}$ and ${\hic\BB}$
and thereby on all fields $\vf$ and antifields $\vf^*$. 
 
The functional \equ{action} or \equ{action-comp} is invariant under
the VSUSY-transformations \equ{vsusy1} or \equ{vsusy2}
up to terms which are linear in the quantum fields \cite{bvwien}
and thus controllable in the quantized theory.
This result also holds after performing the gauge-fixing 
using the BRST operation \equ{svar},
provided specific gauge conditions are considered
  and provided the manifold ${\cal M}_3$ 
admits covariantly constant vector fields $\tau$,
as emphasized in subsection 1.1.1.

\subsubsection{``Diffeomorphisms imply VSUSY"} 

 Let us now show how local VSUSY appears in a natural way, 
within the BRST symmetry, as soon as (external) diffeomorphism
invariance is included in the latter.    
A convenient way to incorporate external diffeomorphisms is to 
consider the extended forms 
$$
\hat{\AA} = \hat c + \hat A + \hat B^* + \hat \phi^*
\quad , \quad
\hat{\BB} =  \hat{\vp} + \hat B + \hat A^* + \hat c^* 
$$
defined by~\cite{exp-xi, bb}  
\begin{equation}
{\hat\AA} = \re ^{-i_{\xi}} {\hic\AA} \quad , \quad 
{\hat\BB} = \re ^{-i_{\xi}} {\hic\BB} \ ,  
\label{ansatz3d}
\end{equation}
 with the extended forms ${\hic\AA}$ and ${\hic\BB}$ given 
 by \equ{antifields3d}.
More explicitly:
\begin{equation}\begin{array}{lcl}
  {\hat\vp}^* = {\hic\vp}^* 
&, \quad & {\hat c}^* = {\hic c}^* 
  \es
   {\hat B}^* = {\hic B}^* - i_{\xi} {\hic\vp}^* 
& , \quad & {\hat A}^* = {\hic A}^* - i_{\xi} {\hic c}^*
  \es
  {\hat A} =\hic A - i_{\xi} {\hic B}^* 
+ {1 \over 2}   i_{\xi}^2 {\hic\vp}^* 
& , \quad &  {\hat B} =\hic B - i_{\xi} {\hic A}^* 
+ {1 \over 2}   i_{\xi}^2 c^* 
  \es
  {\hat c} = {\hic c} - i_{\xi}\hic A 
+ {1 \over 2} i_{\xi}^2 {\hic B}^*  -{1 \over 6}  
 i_{\xi}^3 {\hic\vp}^* 
& , \quad & {\hat \vp}= {\hic\vp} - i_{\xi}\hic B 
+ {1 \over 2} i_{\xi}^2 {\hic A}^*  -{1 \over 6}  
 i_{\xi}^3 {\hic c}^* 
  \ .
   \end{array} 
\label{repa}\end{equation}
By applying the operator $\re ^{-i_{\xi}}$
to the horizontality conditions 
(\ref{gradeeq}) and  using  the last of relations (\ref{ident1}), 
we obtain
\eq\ba{l}
0 = \re ^{- i_{\xi}} \left( {\hic \d} {\hic\AA}+ {\hic\AA} ^2
  + \a {\hic\BB}^2 \right) 
= \left( s + d - {\cal L}_{\xi} + i_v  \right) {\hat \AA} 
+  {\hat \AA} ^2
+ \a {\hat \BB}^2 \es
0 = \re ^{- i_{\xi}}\left( {\hic \d} {\hic\BB} 
+ \left[ {\hic\AA} ,{\hic\BB} \right] \right) 
= \left( s + d - {\cal L}_{\xi} + i_v  \right) {\hat \BB} 
+ \lc {\hat \AA},{\hat \BB} \rc
\, ,
\ea\eqn{hatgradeeq}
where we  have introduced the even (ghost-number $2$) vector field 
\eq
v \equiv s\xi - \xi^2
\qquad 
{\rm with} \ \, 
\xi ^2 \equiv {1\over 2} \, { [ \xi, \xi ] }
\, .
\eqn{def-v}
Equations (\ref{hatgradeeq}) can be rewritten as 
\begin{equation}
{\hat \FF} + \a {\hat \BB} ^2 = 0
\quad , \quad 
{\hat \DD} {\hat \BB} = 0
\ ,
\end{equation}
where 
\begin{equation}
{\hat \FF} \equiv {\hat \d} {\hat \AA}  + {\hat \AA} ^2   
\quad , \quad 
{\hat \DD} {\hat \BB} \equiv {\hat \d} {\hat \BB} +[{\hat \AA}, {\hat \BB} ]	
\ , 
\end{equation}
with~\cite{bb} 
\begin{equation}
{\hat \d} \equiv 
\re ^{-i_{\xi}} \, \delta \, \re ^{i_{\xi}}
= d+s -\LL_{\xi} +i_v 
	\ .
\end{equation}
Hence, they again have the same form as the equations of motion or as the 
horizontality conditions \equ{gradeeq}.  
They determine the action of the BRST operator 
on  the extended forms ${\hat \AA}$ and ${\hat \BB}$:  
\eq\ba{l}
s {\hat \AA} =-  \left( d {\hat \AA} + {\hat \AA}^2 
+ \a {\hat \BB}^2  
- {\cal L}_{\xi} {\hat \AA}
 + i_v  {\hat \AA} \right)  
 \es 
s {\hat \BB} = - \left( d {\hat \BB}
+ \lc {\hat \AA}, {\hat \BB} \rc  
- {\cal L}_{\xi} {\hat \BB}
 + i_v  {\hat \BB} \right)   
 \ .
\ea\eqn{gradebrs}
Thus, they also provide the variations of all component fields, 
\eq
\ba{l} 
 s{\hat c} =  - {\hat c}^2 - \a {\hat \vp}{}^2
  + \LL_\xi{\hat c}
 - i_v \hat A   
 \es
 s{\hat A} = - \hat D {\hat c} - \a \lc {\hat \vp}, {\hat B} \rc  
 + \LL_\xi{\hat A}  - i_v{\hat B}^*  
 \es
 s{\hat B}^* = - ( \hat F + \a {\hat B}{}^2 ) 
 - \lc {\hat c}, {\hat B} ^* \rc 
- \a  \lc {\hat \vp}, {\hat A}^* \rc 
+ \LL_\xi {\hat B}^* 
 - i_v {\hat \vp} ^* 
  \es
 s{\hat\vp}^* = - {\hat D} {\hat B}^* - \lc {\hat c} , {\hat\vp} ^* \rc 
- \a \lp  \lc {\hat\vp}, {\hat c}^* \rc 
 + \lc {\hat B}, {\hat A}^* \rc \rp + \LL_\xi {\hat \vp}^* 
\\[5mm] 
 s{\hat \vp} = - \lc {\hat c}, {\hat \vp} \rc + \LL_\xi{\hat \vp} 
    -i_v {\hat B}   
     \es 
 s{\hat B} = - \hat D{\hat \vp} - \lc {\hat c}, {\hat B} \rc
 + \LL_\xi{\hat B} - i_v {\hat A}^*
  \es 
 s{\hat A}^* = - {\hat D}{\hat B} - \lc {\hat c}, {\hat A}^* \rc 
 -\lc {\hat B}^*, {\hat \vp} \rc + \LL_\xi{\hat A}^* 
    -i_v {\hat c}^*   
      \es
 s{\hat c}^* = - \hat D{\hat A}^* - \lc {\hat c}, {\hat c}^* \rc 
 -\lc {\hat B}^*, {\hat B} \rc 
 -\lc {\hat \vp}^*, {\hat \vp} \rc + \LL_\xi{\hat c}^*  
 \, , 
 \ea
\eqn{hat-brsaf}
where $\hat F \equiv  d {\hat A} + {\hat A}^2$ and 
${\hat D} \cdot \equiv d \cdot + \, [{\hat A}, \cdot \, ]$.  
 Note that the $s$-operator can be decomposed according to 
\begin{equation}
\label{point}
s = s_g + s_{\xi} + s_v 
\ ,
\end{equation}
where the action of $s_g$ and $s_v$ have the form
\equ{brsaf1}\equ{brsaf2} and \equ{vsusy2}, respectively, 
while the action of $s_{\xi}$ is given by the Lie derivative
$\LL_\xi$. 

 So far, the vector field $v \equiv s\xi - \xi^2$
  determining 
  the $s$-variation of $\xi$ was not yet specified.  
By requiring the nilpotency of $s$ on $\hat\AA$,
$\hat\BB$ as well as
 on $\xi$, we do not obtain any constraint on $v$,
except that it must
 transform according to 
\eq
sv = [\xi,v]\ .
\eqn{brs-v}
Henceforth, we conclude that \equ{def-v} (with $v$ subject to the 
transformation law (\ref{brs-v})) 
can
be interpreted as the most general 
BRST-transformation of the diffeomorphism ghost $\xi$ which is 
compatible with nilpotency:
\eq
s\xi = \xi^2 + v
\ .
\eqn{brs-xi}
 We thus see how the local VSUSY-transformations \equ{vsusy2} 
appear in the BRST operator once the diffeomorphisms have been incorporated: 
they are nothing but the symmetry 
transformations of the fields $\hat{\vf}$ and $\hat{\vf} ^*$
whose corresponding ghost is the even vector field $v$ of ghost-number 2. 

To simplify the notation, 
we will drop  
the symbol $\,\hat{}\ $
on fields and extended forms
in the remainder of this subsection.
 An action $S$ which is invariant under the BRST symmetry defined by the 
transformations (\ref{hat-brsaf})-(\ref{brs-xi}) can be constructed 
along the lines followed at the end of subsection \ref{the-model}. 
 The {\em Slavnov-Taylor identity} 
to be satisfied by the action $S$ now takes the extended 
form\footnote{As a matter of fact, a similar Slavnov-Taylor identity
incorporating all of the symmetries had initially been considered 
for $3d$ Chern-Simons theory in flat space, see the second of 
references~\cite{vec-susy}.}
\eq
\SS( S) \equiv  
{1 \over 2} \, 
(S,S) +  \D  S = 0
\ ,
\eqn{slavnov-ext} 
where the BV antibracket $(\cdot,\cdot)$ is given by \equ{BV-bracket}
and where the operator $\D$ is defined by  
\eq 
\D \equiv \dint 
d^3 x 
\lp  s\xi^\m\dfud{}{\xi^\m} + sv^\m\dfud{}{v^\m} \rp
\qquad \mbox{with }
\ \, \D ^2=0 \ . 
\eqn{delta}
The  corresponding linearized Slavnov-Taylor operator reads
\eq
\SS_S \,\cdot \equiv   (S,\,\cdot) + \D\,\cdot  
\ \, , 
\eqn{slavnov-lin-ext}
and we have the following identities
resulting from 
\equ{BV-def-cond} and \equ{special-jacobi}:
\eq 
\ba{l}
\D \SS(S)  = -\SS_S \D S 
\qquad ,\qquad
\SS_S\SS(S) =0
\es
(X, \SS(S)) + (\SS_S)^2 X  =0 
\quad \mbox{for any functional} \ \, X(\vf,\vf^*,\xi,v) 
    \ .
\ea
\eqn{jacobi-ext}
Following the arguments of subsection 2.1, an 
action obeying the Slavnov-Taylor identity
\equ{slavnov-ext}  is found by solving the functional
differential equations 
 \eq\ba{l}
\SS_S\vf 
\equiv 
\dfud{S}{\vf^*} = s\vf 
  \quad ,\quad 
\SS_S\vf^* 
\equiv 
\dfud{S}{\vf} = s\vf^*
\ , 
\ea\eqn{differential-eqs-bis}
 where the BRST operator
$s$ is now given by (\ref{hat-brsaf})-(\ref{brs-xi}). 
The solution represents an extension of
expression \equ{action}, 
\begin{equation}
\label{action-ext}
S = - \int_{{\cal M}_3} \tr \left. \lp 
{\hhat\BB} \lp  d\hhat\AA + \hhat \AA ^2\rp
+ {\a \over 3} 
{\hhat\BB}^3 - {\hhat\BB} 
( \LL_{\xi} {\hhat\AA} - i_v {\hhat\AA} ) 
\rp  \right| _3
\, , 
\end{equation} 
or, in components, 
\eq
\ba{l}
S =  -\dint \tr \lp  \hhat B \hhat F + \dfrac{\a}{3}{\hhat B^3} 
\rp  + 
\dsum {\vf}{} \dint \tr \lp  \vf^* \bar s \vf \rp   
 - \dint \tr \lp  {\hhat {\hic\vp}}^* i_v {\hhat B} + {\hhat c}^* i_v {\hhat A} +
{\hhat A}^* i_v {\hhat B}^*  \rp  \es
\phantom{S} \equiv  S_{\rm inv} ( \hhat A, \hhat B ) 
+ S_{\rm
antifields} ( \vf, \vf^* ) 
\ .
\ea
\eqn{action-comp-ext}
 Here, $\bar s{}{\vf}$ denotes the BRST-variations \equ{hat-brsaf}
taken at $v=0$, i.e. without vector supersymmetry, the effect 
of the latter appearing explicitly in the third integral.

We note that the action involves 
a contribution that is quadratic in the antifields. 
This term reflects the fact that  the algebra of gauge
symmetries, diffeomorphisms and vector supersymmetry 
only closes on-shell, i.e. by virtue of the equations of 
motion~\cite{vec-susy,cs-lucch-pig}.

 To verify that the 
 Slavnov-Taylor identity \equ{slavnov-ext} 
 is satisfied, we again proceed as in   
  subsection \ref{the-model},
  by applying the last of the identities
  \equ{jacobi-ext} and using 
  $\SS_S{}^2=0$. 
  As before, this implies that 
$\SS(S)$ is independent of the fields and antifields $\vf$, 
$\vf^*$. Hence $\SS(S)$ can only depend on the variables 
$\xi$ and $v$,  
\eq
\SS(S) = F(\xi,v) = \dint d^3x\, a_\m\xi^\m\ ,
\eqn{an-slavn}
where we used the fact that 
the functional $F$ has to be 
linear in $\xi$, as well as independent
 of $v$ due to ghost-number
conservation (the coefficient $a_\m$ is field independent). 
The second of the identities \equ{jacobi-ext} then yields the
consistency condition $\D F=0$ whose solution is $a_\m=0$.

The main  conclusions of the present section are the following.
 First, by incorporating 
diffeomorphisms into the BRST operator, we have shown that the
presence of VSUSY as a {\em local} symmetry is natural 
in the sense that it
belongs to the most general BRST algebra (compatible with nilpotency)
for the present set of fields.
 Second, the inclusion of VSUSY-transformations 
into the $s$-variations allows for their discussion 
on {\em generic} manifolds.


\subsection{Correspondence between $3d$ gravity and  
BF models}\label{corresp-gr-bf}

\subsubsection{Action and BRST symmetry}

 We now choose $so(3)$  or $so(1,2)$ as symmetry 
algebra in order to 
establish the relationship between 
the associated BF models
and $3d$ gravity, 
extending off-shell the on-shell result presented in
subsection \ref{grav-bf}.  The
correspondence appeared there via the substitution \equ{change}, in
particular via the reparametrization 
${\hic\vp}$ = $i_{\xi}\hic B$ of the ghost 
$\hic\vp$ in terms of the diffeomorphism ghost $\xi$. 
In view of relations \equ{repa}, 
we infer that the off-shell extension of the former equation
reads
\eq
\hat\vp = 0  
\ , \qquad 
\mbox{i.e.}
\quad  
{\hic\vp}= i_{\xi}\hic B - {1 \over 2} \, i_{\xi}^2  {\hic A}^* 
+ {1 \over 6} \, i_{\xi}^3 {\hic c}^*
\ .
\eqn{change-of-shell}
From the transformation law of $\hat\vp$ 
as given in \equ{hat-brsaf}, we see
that the necessary and sufficient condition for setting $\hat \vp$
to zero  consistently is to set $v$ to zero -- which condition
for its part 
is compatible with the transformation law of $v$, see eq.\equ{brs-v}.
Thus, it is necessary to keep VSUSY out of the BRST operator.

For ${\hat\vp} = 0 =v$,  
the $s$-variations \equ{hat-brsaf} 
and \equ{brs-xi} reduce to 
\eq
\ba{lcl} 
 s{\hat A} = - \hat D {\hat c} + \LL_\xi{\hat A}
 
&,&  s{\hat A}^* = - \hat D{\hat B} - \lc {\hat c}, {\hat A}^* \rc 
  + \LL_\xi{\hat A}^* 
  \es
 s{\hat B} = - \lc {\hat c}, {\hat B} \rc + \LL_\xi{\hat B}
&,& s{\hat B}^* = - ( \hat F + \a {\hat B}{}^2 ) 
 - \lc {\hat c}, {\hat B} ^* \rc 
+ \LL_\xi {\hat B}^*  
  \es 
s{\hat c} =  - {\hat c}^2 + \LL_\xi{\hat c}
&,&  s{\hat c}^* = 
  - \hat D{\hat A}^* - \lc {\hat c}, {\hat c}^* \rc 
 -\lc {\hat B}^*, {\hat B} \rc 
  + \LL_\xi{\hat c}^*  
  \es
s \xi = \xi^2 
  \ . &&
\ea\eqn{3d-grav}
Furthermore, the action  \equ{action-comp-ext} reduces to 
\begin{equation}
\label{comp-ext}
S =  -\dint \tr \lp  \hat B \hat F + \dfrac{\a}{3}{\hat B^3} 
\rp  + 
\dsum {\hat{\vf} =\hat A ,\hat B ,\hat c}{} 
\dint \tr \lp \hat{\vf}^* s \hat{\vf} \rp 
\, ,
\end{equation}
with $s{}{ \hat \vf}$ given by the previous set of equations. 

 With  notation \equ{change} for the Lorentz connection and 
 the 3-bein,
\begin{equation}
   \label{ide}
\hat A_{jk} = \omega_{jk}  \quad , \quad 
\hat c_{jk} = \Omega_{jk} \quad , \quad 
\hat B_{jk} = \varepsilon_{ijk} e^i 
\ , 
\end{equation}
the $s$-variations \equ{3d-grav} become 
  the BRST-transformations of gravity, 
i.e. eqs.\equ{trgr}
and the action \equ{comp-ext} becomes the action 
for gravity involving a cosmological term, i.e. the action based on 
  the invariant functional 
\begin{equation}
\label{sgravc}
S_{\rm inv}\lp e,\omega \rp  = 
- \int  \varepsilon_{ijk} \lp  e^i R^{jk} 
+ \dfrac{\alpha}{3} e^i e^j e^k \rp  
\ .
\end{equation}

 Since diffeomorphisms now represent a dynamical symmetry -- 
the vector field $\xi$ being a dynamical Faddeev-Popov field -- one has
to introduce an antifield\footnote{The $s$-variation of $\xi^*_{\mu}$ 
is determined by 
$s\xi^*_{\mu} \equiv  \d S /\d \xi^{\mu}$.}
$\xi^*$ coupled to the BRST-variation of $\xi$, i.e.
add the term $\int \xi^*_{\mu} s \xi^{\mu}$ to the action 
\equ{comp-ext}. BRST invariance is then expressed by the 
Slavnov-Taylor identity
\equ{slavnov}
in which $\vf$ now takes the values $A$, $e$, $\Omega$ and
$\xi$.

These results generalize the relations found in 
references~\cite{horo,ezawa} 
where the antifields have not been taken into account.

\subsubsection{Vector supersymmetry}\label{vect-supersymm}

 Since we could not include VSUSY-variations into the 
BRST-transformations when considering the gravity theory variables,
we now have to deal with  
this symmetry 
separately, i.e. as an extra symmetry
expressed by a separate Ward identity. 
Thus, we consider the $s$-variations 
  \equ{hat-brsaf}-\equ{brs-xi} with $v=0$ and the VSUSY-transformations 
  \equ{vsusy1}  with an 
   even vector field $\tau$ as parameter. 
Moreover, we assume that ${\hat\phi}=0$
  as in the last section, since this truncation can be performed
  in a consistent way for $v=0$.

 Before evaluating the action of the operator 
 $[\d_{\tau}, s]$ on ${\hat \AA}$
 and  ${\hat \BB}$, we have to specify the VSUSY-variation 
  of $\xi$: in view of the $v$-dependent contribution to $s\xi$ 
  (see eq.\equ{brs-xi}), we postulate 
 \begin{equation} 
\d_{\tau} \xi = k \tau
\, , 
\label{pos} 
  \end{equation}
where  $k$ denotes a  constant. It follows that 
$\left[ \d_{\tau}, i_{\xi}  \right] = k i_{\tau}$ and thereby   
\begin{equation}
\left[ \d_{\tau}, \re ^{-i_{\xi}} \right] = 
- k i_{\tau} \re ^{-i_{\xi}}
\ .
\end{equation} 
By virtue of the definition \equ{ansatz3d}, we have 
\eq\ba{l}
\d_{\tau}{\hat \AA} =
 \d_{\tau} \left( \re ^{-i_{\xi}}\hic\AA\right)  \es
\phantom{\d_{\tau}{\hhat \AA}}
= \left[ \d_{\tau}, \re ^{-i_{\xi}} \right]\hic\AA
+ \re ^{-i_{\xi}}i_{\tau}\hic\AA 
\es
\phantom{\d_{\tau}{\hhat \AA}}
= \left[ \d_{\tau}, \re ^{-i_{\xi}} \right]\hic\AA+ 
\left[ \re ^{-i_{\xi}}, i_{\tau} \right]\hic\AA + i_{\tau} {\hat \AA}
\ . 
\ea\eqn{deltac-hat}
Since the second commutator in  the 
last line vanishes, we obtain 
\begin{equation}
\d_{\tau}{\hat \AA} = ( 1 - k ) i_{\tau} {\hat \AA}
\ ,
  \label{deltaAA-hat1}
\end{equation}
an analogous result holding for ${\hat\BB}$. 
From this relation and eqs.\equ{gradebrs}, \equ{li}, we readily find 
\begin{equation}
\left[ \d_{\tau}, s \right] {\hat \AA} 
= {\cal L}_{\tau}{\hat \AA} - \lp  1 
- k \rp i_{\left[ \xi , \tau \right]} {\hat \AA}
\label{valgebraAA-hat1}
\end{equation}
and analogously for ${\hat\BB}$. 
Choosing $k=1$, we get the $\d_{\tau}$-variations 
\begin{equation}
\label{vsusy3}
\d_{\tau} \xi =  \tau 
\quad , \quad 
\d_{\tau} {\hat\AA} = 0 
\quad , \quad
\d_{\tau} {\hat\BB} = 0 
\ ,
  \end{equation}  
which satisfy the algebra 
\eq
\left[ \d_{\tau}, s \right] = {\cal L}_{\tau}\ ,\quad
s^2 =0
  \quad ,\quad 
    \lc \d_{\tau_1}\, ,  \d_{\tau_2} \rc = 0
    \ .
\eqn{vsusy-alg}
 If we pass over from the BF model to gravity by virtue of the 
identifications \equ{ide}, we recover the VSUSY-variations \equ{vsgr}
of gravity,  i.e. the result of reference \cite{piguet}, now generalized by 
the presence of the antifields. 

 As we already noted at the end of subsection \ref{grav-bf}, the
relation between both versions of $3d$ gravity, namely 
the topological (i.e. BF) version and the conventional one, 
as expressed by eq.\equ{change-of-shell}, is not one-to-one unless the
$3$-bein coefficients (or the
metric) represent a nonsingular matrix.

\subsection{Gauge-fixing}

As usual~\cite{alg-ren}, 
the  theory is gauge-fixed by 
introducing pairs of antighosts and Lagrange multipliers 
associated to
each of the gauge invariances.  We will consider the BF
version of the theory.
Since the  gauge
transformations of $A$ and $B$ 
represent irreducible symmetries in three dimensions, it suffices 
to consider 
one pair of antighosts and multipliers for each of them. 
These pairs of $0$-forms are denoted by 
$\bar c, \pi$
and $\bar\f, \la$, respectively and we have the BRST-transformation
laws 
\eq\ba{lcl}
s\bar c = \pi
\quad & ,&\quad 
s\pi=0
\es
s\bar\f = \la
\quad &,&\quad 
s\la=0
\ .\es
\ea\eqn{antig-lagr}
In the remainder of this subsection, we will again omit the hats on fields
and antifields in order to simplify the notation.
Following the Batalin-Vilkovisky procedure~\cite{bat-vilk}, we first
complete the set of antifields by introducing antifields 
associated to the new
fields, namely the 3-forms ${\bar c}^{\, *}$, $\pi^*$, 
${\bar \f}{}^*$ and $\la^*$.
The latter admit the transformation laws \cite{bvwien} 
\eq\ba{lcl}
s \bar c^{\, *} = 0
  \quad & ,& \quad 
  s \pi^* = \bar c^{\, *} 
\es
s\bar \f^* = 0
  \quad & , &\quad 
  s \la^* =\bar \f^*
    \es
\ea\eqn{antianti}
and enter the so-called {\em non-minimal} action
\eq
S_{\rm nm}(\vf,  \pi, \la
, 
\vf{}^*, \bar c ^{\, *} ,\bar\f{}^* )
=
S(\vf,
\vf{}^*) +\dint \tr 
\lp \bar c ^{\, *} \pi +\bar\f{}^*\la \rp\ ,
\eqn{min-action}
where $S$ is the action \equ{action-comp-ext} or \equ{action-ext}.
This non-minimal 
action solves the same Slavnov-Taylor identity 
\equ{slavnov-ext} as $S$, the summation in \equ{slavnov-lin}
now including the new fields.

Next, we consider the  ``gauge fermion'' func\-tion\-al 
for a Landau gauge,
   
\eq
\Psi(A,B,\bar c,\bar\f) 
= 
\dint \tr
\lp \bar c\, d\star A + \bar\f\, d\star B \rp\ ,
\eqn{g-fermion}
where the Hodge duality operator $\star$ is given by eq.\equ{hodge}.
The fields $\vf, \bar c,\bar\f,\pi,\la$
are to be denoted collectively by $\Phi$ and the corresponding antifields 
by $\Phi ^*$. 
We redefine the antifields according to 
\eq
\check{\Phi} ^* = \Phi^* + \dfud{\Psi}{\Phi}\ ,
\eqn{red-antif-ger}
where $\check{\Phi} ^*$ is to be viewed as the {\em external source} 
associated to
the BRST variation of the field $\Phi$.
The non-trivial reparametrizations read as  
\eq\ba{lcl}
\check A{}^* = A^* + \star \, d\bar c
& \quad ,\quad
& 
\check{\bar c}^{\, *} = \bar c^{\, *} + d\star \! A\ 
\es
\check B{}^* = B^* + \star \, d\bar\f
& \quad ,\quad
& 
\check{\bar\f}{}^* = \bar \f^* + d\star \! B\ .
\ea\eqn{red-antif}

According to the BV prescription, the 
{\em gauge-fixed action} is given by the non-minimal action 
\equ{min-action} 
in which the antifields are reparametrized in terms of the sources 
$\check{\Phi}^*$ by virtue of eq.\equ{red-antif-ger}:
\eq
S_{\rm gauge-fixed} ( \Phi, \check{\Phi}^* ) 
= S_{\rm nm} ( \Phi, \Phi ^* = \check{\Phi} ^* -  \dfud{\Psi}{\Phi} )
\ .
\eqn{g-fixed-action}
Thus, we obtain
\eq
S_{\rm gauge-fixed}  = S_{\rm inv} + S_{\rm gf} + S_{\rm ext} 
+ S_{\rm quadr}
\ ,
\eqn{gaf}
with 
\begin{eqnarray}
&&S_{\rm gf} = - \int  {\rm Tr}
\left( \pi \, d\star \! {\hhat A}  + \lambda \, d\star \! {\hhat B} 
- \bar c \, d \star \! \bar s A 
- \bar{\f} \, d \star \! \bar s B
- (\star \, d\bar c ) \, i_v (\star \, d\bar{\f} )
\right)
\nonumber \es
&& 
S_{\rm ext} = \sum_{\Phi} 
\int  {\rm Tr} \left( 
\check{\Phi} ^*  {\bar s} \Phi  \right)
 - \int  {\rm Tr} \left( 
\check{\phi} ^*  i_v B  
+ \check{c} ^*  i_v A
+ \check{B} ^*  i_v (\star \, d\bar c )
+ \check{A} ^*  i_v (\star \, d\bar{\f} )
\right)
\nonumber \es
&&S_{\rm quadr} =  - \int  {\rm Tr} \left( 
\check A ^* i_v \check B ^* 
\right)
\ , 
\nonumber
\end{eqnarray}
where $\bar s$ is the BRST operator \equ{hat-brsaf} taken at $v=0$.
This action obviously fulfills the same Slavnov-Taylor identity 
as the
non-minimal action.

We note that VSUSY 
continues to hold as a {\it local} invariance of the gauge-fixed theory.
This is in contrast to the results obtained by 
alternative implementations of the 
gauge-fixing procedure~\cite{cs-lucch-pig} for which   
VSUSY only holds as a rigid invariance, the possible vector
fields $v$ being
 covariantly constant (assuming that such vector
fields exist on the manifold under consideration). 
 The more general validity  of
VSUSY invariance in our approach
 is due to its inclusion
in the Slavnov-Taylor identity whereas 
it is left
outside 
in the other schemes.
As for the usual gravitational formulation
discussed in section \ref{corresp-gr-bf}, one is precisely in the latter situation
since we were forced to keep VSUSY outside of the BRST
operator: VSUSY then represents a rigid symmetry, 
$v$ being a Killing vector field with respect to a background 
metric~\cite{piguet}
(assuming again that the 
manifold under consideration admits such vector fields at all).

  \subsection{Summary}

Let us briefly summarize our approach and results. 
\begin{itemize}
\item {\em BF model without diffeomorphisms:} As is well known, the  
gauge transformations of the BF model can be obtained from the 
horizontality conditions ${{\hic\FF}} + \alpha {\hic\BB}^2 =0
= {\hic\DD} {\hic\BB}$
which are equivalent to the 
$s$-variations of ${\hic\AA}$ and $\hic\BB$ given by eqs.\equ{svar}.  
\item {\em Local VSUSY of the BF model on a generic manifold
(or inclusion of diffeomorphisms and VSUSY):} 
By applying the operator ${\rm e}^{-i_{\xi}}$ to the previous 
horizontality conditions and defining $v \equiv s\xi - \xi^2$ 
(i.e. $s\xi = \xi^2 +v$), 
one can derive the $s$-variations of
the reparametrized fields ${\hat\AA} \equiv {\rm e}^{-i_{\xi}} \hic\AA, \, 
{\hat\BB} \equiv {\rm e}^{-i_{\xi}} {\hic\BB}$, see eqs.\equ{gradebrs},
 which include diffeomorphisms and local VSUSY. 
\item {\em From the BF model to $3d$ gravity:}
We set $v=0$ (in order to eliminate VSUSY from the $s$-variations)
and ${\hat\phi} =0$ (in order to express the gauge transformation
of the $1$-form potential $B$ in terms of diffeomorphisms). 
By virtue of the identification \equ{ide}, 
the invariant action of the BF model and its local symmetries
then yield those of gravity. 
The VSUSY-transformations of gravity 
are recovered {\em directly} from the VSUSY-variations of the BF model
as given by eqs.\equ{vsusy1}, by virtue of the 
reparametrization ${\hat\AA} \equiv {\rm e}^{-i_{\xi}} \hic\AA, \, 
{\hat\BB} \equiv {\rm e}^{-i_{\xi}} {\hic\BB}$, the identification \equ{ide} 
and the postulate $\d_{\tau} \xi = \tau$. 
\item  {\em  Gauge-fixing:} Applying the Batalin-Vilkovisky gauge-fixing 
procedure to the BF theory, with VSUSY included into the 
BRST-transformations, we have seen that
this 
symmetry holds  as a local invariance of the action. 
Yet, this result is no longer valid 
if VSUSY is kept out of the BRST operator, as expected 
for the conventional formulation of $3d$ gravity.
\end{itemize}

\section{4d BF model with cosmological term}

\subsection{The model and its symmetries} 
   
The model is now defined on a $4$-manifold ${\cal M}_4$ 
and the potential ${\hic B}$ represents a $2-$form.
The action reads 
\begin{equation}
S_{\rm inv} \lp {\hic A},{\hic B}\rp  = - \int_{{\cal M}_4} 
\tr \lp  {\hic B}{\hic F} + \frac{\la}{2} {\hic B}^2 \rp  
\, , 
\label{bf4d}
\end{equation}
where the real dimensionless parameter $\lambda$ is again referred to as
cosmological constant. 
The equations of motion of this model have the form 
\begin{equation}
{\hic F} + \la {\hic B} = 0
\quad , \quad {\hic D}{\hic B} = 0 
\ . 
\label{eom4d}
\end{equation}
In this case, we consider the generalized fields 
\begin{eqnarray}
{\hic\AA}  \!\!\!&=&\!\!\! {\hic c} + {\hic A} + {\hic B}^* + {\hic B}_1 ^* + {\hic B}_0 ^*
\qquad ( \, {\rm with} \ \; {\hic B}_i ^* \equiv ({\hic B}_i) ^* \, )
\nonumber 
\es
{\hic\BB} \!\!\!&=&\!\!\! {\hic B}_0 + {\hic B}_1 + {\hic B} 
+ {\hic A}^* + {\hic c}^*
\, ,
\label{antifields4d}
\end{eqnarray}
where ${\hic B}_1$ denotes the ghost parametrizing the local gauge symmetry 
of ${\hic B}$ while ${\hic B}_0$ represents the ``ghost for the ghost" ${\hic B}_1$.

The horizontality conditions have the same form as the equations 
of motion, 
\begin{eqnarray}
{{\hic \FF}} + \lambda {\hic \BB} = 0 
\qquad , \qquad 
{{\hic \DD}} {\hic \BB} = 0
\label{hc}
\end{eqnarray}
and define a nilpotent
BRST operator.
Expansion with respect to the ghost-number yields the BRST-transformations
of fields and antifields: 
\begin{eqnarray}
s{\hic c} \!\!\! &=& \!\!\! - {\hic c}^2 - \la {\hic B}_0  
\nonumber \es
s{\hic A} \!\!\! &=& \!\!\! - {\hic D} {\hic c} - \la {\hic B}_1   
 \nonumber \es
s{\hic B}^* \!\!\! &=& \!\!\! - ( {\hic F} + \la {\hic B} )  - \lc {\hic c}, {\hic B} ^* \rc 
 \nonumber \es
s{\hic B}_1^* \!\!\! &=& \!\!\! 
 -{\hic D}{\hic B}^* - \lc {\hic c}, {\hic B}_1^* \rc - \la {\hic A}^* 
 \nonumber \es
s{\hic B}_0^* \!\!\!&=&\!\!\! - {\hic D}{\hic B}_1^* - ( {\hic B}^*)^2  -
\lc {\hic c} , {\hic B}_0^* \rc - \la {\hic c}^*  
\label{brsaf1-4d}
\end{eqnarray}
and 
\begin{eqnarray}
s{\hic B}_0 \!\!\! &=& \!\!\! -\lc {\hic c}, {\hic B}_0 \rc  
\nonumber \es
s{\hic B}_1 \!\!\!&=&\!\!\! - {\hic D}{\hic B}_0 - \lc {\hic c}, {\hic B}_1 \rc  
\nonumber \es 
s{\hic B} \!\!\!&=&\!\!\! - {\hic D}{\hic B}_1
 - \lc {\hic c}, {\hic B} \rc +  \lc {\hic B}^*, {\hic B}_0 \rc  
\nonumber \es
s{\hic A}^* \!\!\!&=&\!\!\! - {\hic D}{\hic B} - \lc {\hic c}, {\hic A}^* \rc
 - \lc {\hic B}^*, {\hic B}_1 \rc 
+ \lc {\hic B}_1^*, {\hic B}_0 \rc 
\nonumber \es
s{\hic c}^* \!\!\!&=&\!\!\! - {\hic D}{\hic A}^* 
- \lc {\hic c}, {\hic c}^* \rc - \lc {\hic B}^*, {\hic B} \rc 
 -\lc {\hic B}_1^*, {\hic B}_1 \rc  - \lc {\hic B}_0^* , {\hic B}_0 \rc 
\ . 
\label{brsaf2-4d}
\end{eqnarray}
If the antifields are set to zero, we recover 
the BRST-transformations of ${\hic A}$,
${\hic B}$ and of their ghosts, as well as their equations of motion.

\subsection{Diffeomorphisms and vector supersymmetry}

Proceeding along the lines of the  three dimensional theory,
one introduces new generalized fields 
${\hat { \AA}}= \re ^{-i _\xi}{\hic \AA}$ and  
${\hat {\BB}}= \re ^{-i _\xi}{{\hic\BB}}$
as in eqs.(\ref{ansatz3d}),  
which define 
new fields $\hat \vf$ and antifields $\hat \vf^*$.
 
Applying the operator $\re ^{-i _\xi}$ to the horizontality 
conditions (\ref{hc}), we obtain
the $s$-variations of the new fields: 
\begin{eqnarray} 
   s \hat c \!\!\! & = &  \!\!\!
   - \hat c^2 - \lambda \hat B _0 + {\cal L} _{\xi} \hat c 
   - i_v \hat A 
   \nonumber \es
   s \hat A \!\!\! & = &  \!\!\! 
   - \hat D \hat c - \lambda \hat B _1 
   + {\cal L} _{\xi} \hat A - i_v \hat B ^* 
 \nonumber \es 
s {\hat B} ^*  \!\!\! &=& \!\!\!
- ({\hat F}+ \lambda {\hat B})
-[{\hat c},{\hat B}^*] +{\cal L}_{\xi} {\hat B}^*
-i_v {\hat B}^*_1
\nonumber \es
s{\hat B}^*_1 \!\!\! &=& \!\!\!
-{\hat D}{\hat B}^*-[{\hat c},{\hat B}^*_1]-\lambda {\hat A}^*
+{\cal L}_{\xi} {\hat B}^*_1 - i_v {\hat B}^*_0
\nonumber \es
s{\hat B}^*_0  \!\!\! &=& \!\!\!
- {\hat D} {\hat B}^*_1 - \lc {\hat c},{\hat B}^*_0 \rc 
-({\hat B}^*)^2-\lambda {\hat c}^*+{\cal L}_{\xi} {\hat B}^* _0
\nonumber 
\end{eqnarray}
and 
\begin{eqnarray} 
   s \hat B_0  \!\!\! &=& \!\!\!
   - \lc \hat c , \hat B_0 \rc + {\cal L}_{\xi} \hat B_0 
   - i_v B_1 
   \nonumber \es
   s \hat B_1 \!\!\! &=& \!\!\!
- \hat D \hat B _0 - \lc  \hat c , \hat B_1 \rc  
+ {\cal L}_{\xi} \hat B_1 - i_v \hat B 
\nonumber \es
s{\hat B} \!\!\! &=& \!\!\!
-{\hat D}{\hat B} _1 - \lc {\hat c} , {\hat B} \rc 
- \lc {\hat B} ^* , {\hat B}_0 \rc  + {\cal L}_{\xi} {\hat B}
- i_v {\hat A}^*
\nonumber \es
s \hat A^* \!\!\! &=& \!\!\!
- \hat D \hat B - \lc \hat c , \hat A ^* \rc 
- \lc \hat B ^* , \hat B _1 \rc 
- \lc \hat B ^* _1 , \hat B _0 \rc 
+ {\cal L} _{\xi} \hat A ^* - i_v \hat c ^* 
\nonumber \es
s{\hat c} ^* \!\!\! &=& \!\!\!
-{\hat D} {\hat A} ^*- \lc {\hat c} , {\hat c}^* \rc 
-\lc {\hat B}^*,{\hat B} \rc - \lc {\hat B}^{*}_{1},{\hat B}_1 \rc  
- \lc {\hat B}^{*}_{0},{\hat B}_0 \rc +{\cal L}_{\xi} {\hat c}^*
\ , 
\nonumber 
\end{eqnarray}
where ${\hat F}=d{\hat A}+{\hat A}^2$ and 
${\hat D}\cdot =d \cdot +[{\hat A},\cdot ]$.  The BRST-transformations of the
diffeomorphism and supersymmetry ghosts 
again have the form 
\[
s\xi=\xi^2+v
\quad , \quad 
sv=[\xi,v]
\ ,
\]
where the latter relation follows by requiring nilpotency
of $s$ on $\xi$.
\subsubsection*{Slavnov-Taylor identity:}
In the following, we will again omit the hats on fields and antifields.
An action which is invariant under the latter BRST-transformations
that involve VSUSY 
 can be constructed along the lines of  section \ref{BF-model-3d}:
  this action satisfies the Slavnov-Taylor identity \equ{slavnov-ext}
in which the integration is now performed over the $4$-manifold 
 ${\cal M}_4$ and it is explicitly given by 
\eq\ba{ll}
S =\! \! & - \int_{\MM _4} {\rm Tr}
\lp  {\hhat B}{\hhat F} + \frac{\lambda}{2}{\hhat B}^2 \rp  
+\sum_{\vf} \int_{\MM _4} {\rm Tr} \lp  {\vf}^* 
{\bar s}{\hhat \vf} \rp  
\es
  & \qquad \qquad \qquad \qquad \qquad \quad  
  - \int_{\MM _4} {\rm Tr} \lp   
{\hhat A}^* i_v {\hhat B}^* + ({\hhat B}^*)^2{\hhat B}_0  \rp 
\ , 
\ea\eqn{4-d-action}
where ${\bar s}\hhat \vf$ is the part of  $s\vf$
 that does not depend on antifields.
The antifield-dependent part of $s\vf$
is
taken into account by the third integral
containing terms that are quadratic in the antifields. 
As we can see,
a quadratic term of the form  $({\hhat B}^*)^2{\hhat B}_0$
is present even if VSUSY is excluded from the BRST operator, 
as usual for BF models 
in dimensions greater than three \cite{bvwien}. 

\subsection{Gauge-fixed action}

We shall introduce the gauge-fixing term in the action following the 
Batalin-Vil\-ko\-vis\-ky (BV) procedure  as in section 
\ref{BF-model-3d}. By constructing the BV 
pyramid~\cite{bat-vilk, birm-et-al, lps}, 
it can easily be seen that  it is necessary to introduce a  set of 
Lagrange multipliers ($\pi^0_1$, $\pi^{-1}_0$, $\pi^1_0$) and 
antighosts 
$( {\bar c} ^{\, -1}_1 ,
{\bar c} ^{\, -2}_0 ,
{\bar c} ^{\, 0}_0 )
\equiv 
( {\bar c}_1 ,  {\bar \gamma} , {\bar \phi} )$ 
for the reducible gauge symmetry of $B$. 
We also have to 
introduce a Lagrange multiplier $\pi$ 
and an antighost ${\bar c}$ for the gauge symmetry of $A$. 
These fields transform as
\eq
s{\bar c}=\pi 
\quad  ,\quad
s{\bar c}_1=\pi^0_1 
\quad  ,\quad
s{\bar \gamma}=\pi^{-1}_0 
\quad ,\quad
s{\bar \phi}=\pi^1_0
\ ,
\eqn{6.1}
all Lagrange multipliers being $s$-inert. As before, an antifield
is introduced 
for each Lagrange multiplier and each antighost. 
Moreover, all fields are collectively denoted by $\Phi$ and the 
corresponding antifields by $\Phi^*$. 

The non-minimal action reads 
\eq
S_{\rm nm}(\Phi
, \Phi ^* )
=
S(\vf,
\vf{}^*) 
+\dint_{\MM _4}  \tr 
\lp 
\bar c ^{\, *} \pi 
+ \bar c _1 ^{\, *} \pi_1^0  
+ \bar{\gamma} ^* \pi_0 ^{-1}  
+ \bar\f{}^* \pi_0^1 
\rp
\eqn{nm4}
 and as ``gauge fermion'' we choose the functional~\cite{lps, bvwien} 
\eq
\Psi (\Phi) =  \int_{\MM _4}  {\rm Tr} \lp  (d{\bar c}) \star {\hhat A} 
+ (d{\bar c}_1) \star {\hhat B} +
(d{\bar \gamma}) \star {\hhat B}_1 + (d{\bar \phi}) \star {\bar c}_1 \rp \ .
\eqn{6.3}
An external source is associated to each field $\Phi$ 
by virtue of the reparametrization \equ{red-antif-ger} and 
the gauge-fixed action is obtained from the non-minimal action
by virtue of the prescription \equ{g-fixed-action}.
The result again has the form \equ{gaf} and
explicit expressions for all contributions 
can readily be obtained from  
\equ{nm4} and \equ{6.3}.

The generalization to any dimension $d\geq 5$
can be achieved in a straightforward way by introducing the appropriate 
ghosts for ghosts, antighosts and Lagrange multipliers.

\section{ Conclusions}

Our first two conclusions concern VSUSY of BF models. 
First, we are naturally led to the existence 
of a {\it local} vector supersymmetry for BF models 
within the BV framework
if we use the formalism of
extended differential forms and if we include diffeomorphism
 symmetry into the BRST-transformations.  
Second, VSUSY  introduced in this way is 
valid on generic manifolds and still holds exactly as a local 
invariance after carrying out the gauge-fixing procedure. 
These results which we discussed in $3$ and $4$ 
space-time dimensions obviously 
extend to BF models in higher dimensions.
They generalize in a substantial way
the previous results~\cite{vec-susy,cs-lucch-pig}
using an approach for 
which VSUSY only holds, after gauge-fixing, as a rigid symmetry 
generated by covariantly
constant vector fields. 
In fact, the latter approach is restricted to manifolds
admitting such covariantly constant vectors.

Our further conclusions concerning $3d$ field theories are as follows. 
For three dimensions and specific gauge groups,  the BF model 
can be related directly to gravity.
Thus, we could show that 
 VSUSY still holds, as
expected~\cite{piguet}, within the gravity framework.
However, VSUSY  then
turns out to be excluded from the BRST symmetry corresponding to the
invariances with respect to 
diffeomorphisms and local Lorentz transformations. 
Together with the BRST operator $s$, it still obeys the 
usual~\cite{vec-susy,cs-lucch-pig,piguet} algebra \equ{vsusy-alg}. 
As a consequence, after carrying out the gauge-fixing procedure, 
VSUSY can only hold as a rigid
symmetry generated by a
 Killing vector field, as shown in
reference~\cite{piguet}. 
Thus, for $3d$ gravity, 
VSUSY can only be included into the BRST operator 
if the theory is described 
in the topological framework, i.e. to a BF model. 
Of course, 
this conclusion only applies in the $3d$ case, 
since higher dimensional gravity is not topological.
Rather it is related to a 
{\em constrained} BF model whose investigation deserves a separate 
study \cite{ip}.

\section*{Appendix: Notation and Useful Formulas}

\renewcommand{\theequation}{\Alph{section}.\arabic{equation}} 
\setcounter{section}{1}
\setcounter{equation}{0}

All fields that we consider 
are vector fields or Lie algebra-valued 
$p$-forms on a 
$d$-dimensional manifold $\MM_d$ (see the beginning
of subsection \ref{local-vsusy}).
 In the sequel, 
  we will summarize our notation concerning 
 all of these fields and functionals 
  thereof. 
  
\subsubsection*{Differential forms and grading}

The {\it total degree} of a Lie algebra-valued 
  $p$-form $\om_p^g$ of ghost-number $g$
is defined by 
\eq
[\om_p^g] 
= p+g\ . 
\eqn{tot-deg}
If the total degree is even (odd), the form is said to
be {\em even (odd)} and its {\em grading} function
\eq
\mbox{Grading}\,(X) = (-1)^{[X]}
\eqn{grading} 
then takes the values $+1$ and $-1$, respectively. 
For instance, the gauge connection $A$ is odd, since it is 
 a $1$-form with ghost-number $0$ and the Faddeev-Popov ghost $c$ is
odd too, since it is  a
$0$-form with ghost-number $1$. 

The commutator of Lie algebra-valued forms is 
{\em graded} by the total degree, 
i.e. 
\eq
[X,Y] = XY - (-1)^{[X][Y]} YX \ .
\eqn{graded comm}
Thus, the graded commutator of $X$ and 
$Y$ amounts to an anticommutator if both quantities are odd and  
to a commutator otherwise.

The {\it exterior derivative} $d$, which acts as an antiderivation
increasing the form degree by one unit, 
  is defined in local coordinates by
\eq
d= dx^\m \pa_\m\ .
\eqn{ext-dr}
E.g. for the Faddeev-Popov ghost $c$ (which is of ghost-number 1),
we have 
$dc=dx^\m\pa_\m c$ = $-\pa_\m c\,dx^\m$.
  
The {\em BRST differential} $s$ also acts as an antiderivation
  which increases the ghost-number (and thus the total degree)
    by one unit.
A linear operator acting on products like a 
derivation (antiderivation) is 
called an {\em even (odd)} operator.
  The commutator of two such operators is always assumed to 
  be graded according to \equ{graded comm},
 e.g.  $[s,d] =sd +ds$.

The {\it Hodge dual} of a $p$-form $\om$ is 
  the $(d-p)$-form $\star \, \om$ defined by
\cite{bertlmann} 
\eq\ba{ll}
& \star \, \om  = \dfrac{1}{(d-p)!}\, \tilde\om{}_{\m_1...\m_{d-p}}
 dx^{\m_1}... dx^{\m_{d-p}} \es
{\rm where} 
\quad &
\tilde\om{}_{\m_1 ... \m_{d-p}} = \dfrac{1}{p!}
\, \e_{\m_1 ... \m_d} \om^{\m_{d-p+1}...\m_d}
\ .
\ea\eqn{hodge}
  Here and elsewhere in the text, the wedge product symbol 
has been  omitted. Moreover, 
a background metric $(g_{\m\n})$ has been used on ${\cal M}_d$, 
as well as the
totally antisymmetric tensor of Levi-Civita:
\eq
\e_{\m_1 ... \m_d} = 
g_{\m_1\n_1} \cdots
g_{\m_d\n_d}\e^{\n_1 ... \n_d}
\quad ,
\quad 
\e^{1... d}=1
\quad ,\quad 
\e_{1... d} = \mbox{det}\,(g_{\m\n})
\ .
\eqn{lev-citta}
The following formulas are
quite 
useful~\cite{wien}:  
\eq
\star \star \om_p^g = (-1)^{p(d-p)} \, \mbox{det}\,(g_{\m\n})   
\, \om_p^g\ 
\quad , \quad
\tr \! \left( \omega^g_p \star \phi^h_p
\right)
= (-1)^{(p+d) (g+h)+gh} 
\, \tr \! \left( \phi^h_p \star \omega^g_p
\right).
\eqn{dual-comm}
Since the Hodge star operator maps a form of total degree $p+g$ to
a form of total degree $(d-p)+g$, it represents an even operator
  if the space-time dimension is even and an odd operator
otherwise.

  \subsubsection*{Vector fields, inner product and Lie derivative} 
For a vector field $w=w^\m \pa_\m$ 
  on $\MM_d$, 
the total degree is given by its ghost-number $g$.
It is said to be {\em even (odd)} if $g$ is even (odd). 

The Lie bracket $[u,v]$ of two vector fields $u$ and $v$  
is again a vector field: this bracket is assumed to be 
graded according to \equ{graded comm} so that its 
components are  given by 
\[
[u,v]^\m = u^\n\pa_\n v^\m  -(-1)^{[u][v]}   v^\n\pa_\n u^\m 
\ . 
\]

    The  {\em interior product} 
 $i_w$ with respect to the vector field
  $w= w^\m \pa_\m$ is defined in local coordinates by
\eq
i_w\vf = 0 \quad \mbox{ for 0-forms} \ \vf \quad ,\quad
i_w (dx^\m ) = w^\m
\ .
\eqn{int-der}
  If $w$ is even, the operator  $i_w$  acts as an antiderivation 
(odd operator), otherwise it acts 
as a derivation (even operator).

The {\em Lie derivative} $\LL_w$
with respect to $w$ 
acts on differential forms according to 
\eq
\LL_w \equiv [i_w,d] = i_w d  +(-1)^{[w]}  d i_w 
\eqn{Lie-deriv}
and we have the graded commutation relations
\begin{equation}
\lc \LL_u , i_v \rc  =  i_{[u,v]}
\quad  ,\quad 
 \lc \LL_u,\LL_v \rc = \LL_{[u,v]} \ . 
\label{li}
\end{equation}

In the main body of the text, the quantity $\xi = \xi^{\mu} \pa_{\mu}$ 
always denotes a vector
field of ghost-number $1$ (representing the ghost for
diffeomorphisms). We then have  
the following identities involving the vector fields 
$\xi$ and  
$\xi^2 \equiv {1 \over 2} [ \xi , \xi ]$ as well as  
the previously introduced
operators:
\eq\ba{l}
\re ^{i_{\xi}} ( X \, Y )  =  \lp  \re ^{i_{\xi}} X\rp\lp  \re
^{i_{\xi}}Y\rp
\es 
{\rm e}^{- i_{\xi}} d \re ^{i_{\xi}}  =  d - {\cal L}_{\xi} - i_{\xi^2}
\es
\left[ s, \re ^{i_{\xi}} \right]  
=  i_{s\xi}\, \re ^{i_{\xi}} 
\quad  , \quad \left[ s, \re ^{- i_{\xi}} \right]  
= - i_{s\xi}\, \re ^{- i_{\xi}} 
\ .
\ea\eqn{ident1}

\subsubsection*{Functional calculus with differential forms}

The {\it functional derivative} $\d F/\d\om$
of a functional $F$ depending on differential forms $\om$,...
is defined as a left derivative by 
\eq
\d F = \int \d\om \, \dfud{F}{\om}\ ,
\eqn{funct-der}
where $\d F$ is the variation of $F$ induced by the variation
$\d\om$.
In particular, for a $p$-form $\om_p$, 
we have 
\eq
\dfud{\om_p(x)}{\om_p(y)} = \d_{d-p,p}(y,x)\ ,
\eqn{f-der-om}
where the right-hand side is a Dirac-type distribution defined by
\eq
\dint_{y\in \MM_d} 
\om_p(y) \d_{d-p,p}(y,x) = \om_p(x)\ .
\eqn{dirac-dis}

  In order to discuss the grading of $\d F/\d\om$, we first 
recall  that the integral of a $d$-form
over $\MM_d$
is defined by 
\eq
\dint_{\MM_d} \om_d^g = \dfrac{1}{d!}
 \dint_{\MM_d} d^dx\, \e^{\m_1...\m_d} \om^g_{\m_1...\m_d} \ ,
\eqn{def-integral}
where the right-hand side represents the integral written in local coordinates. 
Thus, for an {\em odd} dimension $d$, this prescription 
changes the grading of the
integrand.
For instance, if the integral is an action 
functional, then $g=0$
and
 the integrand is of odd degree, whereas the integral is of even degree.
This fact implies that the integration operator itself has a grading 
and therefore
the grading of a functional derivative also depends on the 
space-time dimension:
\eq\ba{ll}
\mbox{Grading of integral symbol\ }\dint_{\MM_d} :& (-1)^d \es
\mbox{Grading of functional derivative\ } 
       \dfud{F}{\om} :\ &(-1)^{d+[F]+[\om]}\ .
\ea\eqn{gradings}

\subsubsection*{Batalin-Vilkovisky Algebra}
In the following, 
we adapt the formalism of Batalin and Vilkovisky 
(BV)~\cite{bat-vilk} to the case 
where the fields $\vf$ are differential forms on a $d$-dimensional manifold.
The antifield associated to the field $\vf$ is denoted by $\vf^*$, 
the total degrees of the fields and antifields being related by 
\eq
[\vf]+[\vf^*] = d-1
\ .
\eqn{grad-fi-fi*}

 With derivatives operating {\it from
the left}, the BV antibracket of two functionals $X$ and $Y$
depending on $\vf$ and $\vf^*$ 
is defined by 
\eq
(X,Y) = \dsum{\vf}{} \dint _{\MM_d}
\lp (-1)^{[X][\vf^*]} \dfud{X}{\vf^*}\dfud{Y}{\vf} 
  + (-1)^{[X][\vf]+d([\vf]+1)} \dfud{X}{\vf}\dfud{Y}{\vf^*} \rp
  \ .
\eqn{BV-bracket}
Here, the summation is performed on all fields (forms) $\vf$ 
and  $[Q]$ denotes the total degree 
of the quantity $Q$.  
For $Y = \vf$ and $Y= \vf^*$, we obtain respectively 
\eq
(X,\vf) = (-1)^{[X][\vf^*]} \dfud{X}{\vf^*}
\quad , \quad
(X,\vf^*) = (-1)^{[X][\vf]+d([\vf]+1)} \dfud{X}{\vf}
\ .
\eqn{Bf-vphi}
Our definition of the bracket $(X,Y)$ corresponds to the 
 normalization re\-qui\-re\-ments
\eq\ba{l}
\half(S,S) = \dsum{\vf}{} \dint \dfud{X}{\vf^*}\dfud{Y}{\vf}\ 
\es
\half \pa_\la (S,S) = (\pa_\la S,S)\ ,
\ea\eqn{BV-def-cond}
where $S$ denotes a functional of even total degree and $\la$ an even or
odd parameter.

Keeping in mind our definitions 
\equ{funct-der} and \equ{grading} of the functional derivative 
and grading, 
as well as the properties \equ{gradings}, we can derive the
graded anti-symmetry property
\eq
(X,Y) + (-1)^{([X]+1)([Y]+1)}(Y,X)=0
\eqn{BV-symm}
and the graded Jacobi identity
\eq
(X,(Y,Z)) + (-1)^{([X]+1)([Y]+[Z])}(Y,(Z,X))
  + (-1)^{([Z]+1)([X]+[Y])}(Z,(X,Y)) = 0\ .
\eqn{BV-jacobi}
The latter results from the following simple identity which is valid 
  for any 
functional $S$
 of even total degree:
\eq
(S,(S,S)) =0\ .
\eqn{BV-jacobi-S}
Indeed, it suffices to consider $S=xX+yY+zZ$ where the gradings of the
coefficients $x,\,y$ and $z$ are equal to those of $X,\,Y$ and $Z$, 
respectively, then
to differentiate with $\pa^3/\pa x\,\pa y\,\pa z$ at $x=y=z=0$ while 
using the differentiation formula
\eq
\pa_\la(X,Y) = (\pa_\la X,Y) + (-1)^{[\la]([X]+1)} (X,\pa_\la Y)\ .
\eqn{bv-diff}
A useful special case of the Jacobi identity is 
\eq
\left( X,(S,S) \right) + 2 \left( S,(S,X) \right) = 0 
\ ,
\eqn{special-jacobi}
where the total degree of $X$ is  arbitrary.


\end{document}